\documentstyle[12pt]{article}
\newcommand{\nc}{\newcommand}
\nc{\be}{\begin{equation}}
\nc{\ee}{\end{equation}}
\nc{\bea}{\begin{eqnarray}}
\nc{\eea}{\end{eqnarray}}
\nc{\bd}{\begin{displaymath}}
\nc{\ed}{\end{displaymath}}
\nc{\s}{\sum}
\nc{\vi}{Virasoro}
\nc{\wh}{\widehat}
\nc{\ra}{\rightarrow}
\nc{\pa}{\partial}
\nc{\dv}{$D$--\vi}
\nc{\al}{\alpha}
\nc{\til}[1]{\tilde{#1}}
\nc{\lb}[1]{\label{eqn:#1}}
\nc{\rf}[1]{\ref{eqn:#1}}

\begin{document}
\begin{titlepage}
\rightline{TMUP-HEL-9302}
\rightline{March 1993}
\vskip 1.5cm

\begin{center}
\begin{Large}
{\bf  Discretization of Virasoro Algebra}
\end{Large}

\vskip 2cm

Ryuji K{\footnotesize{EMMOKU}}\footnote[2]{E-mail:kemmoku@phys.metro-u.ac.jp}
and Satoru
S{\footnotesize{AITO}}\footnote[3]
{E-mail:saito@phys.metro-u.ac.jp}\footnote[4]
{This work is supported in part by
the Grant-in-Aid for general Scientific Research from the Ministry of
Education, Science and Culture, Japan (No.02640234)}
\vskip  .5cm

{\it Department of Physics, Tokyo Metropolitan University}

{\it Minami-Osawa, Hachioji, Tokyo 192-03, Japan}

\end{center}

\vskip 8cm
\centerline{\bf Abstract}

\vskip .5cm

A $q$-discretization of \vi\ algebra is studied which reduces to the ordinary
\vi\ algebra in the limit of $q \ra 1$.
This is derived starting from the Moyal bracket algebra, hence is a kind of
quantum deformation different from the quantum groups.
Representation of this new algebra by using $q$-parametrized free fields
is also given.

\end{titlepage}
\newpage
\section{Introduction}

 In recent years, one of the main themes of the study of nonlinear integrable
systems is to make clear the physical and mathematical meaning of quantum
groups \cite{inf}.
Physically, quantum groups have been regarded in some sense as hidden
symmetries of
the integrable systems (strings, conformal fields, statistical models, etc.).
Many people discussed deformation of such models as
extension or generalization of these models.
Mathematically, the representation theory of quantum groups has been
investigated based on the results of Lie algebra.
Especially, theories based on $U_{q}(\widehat{sl_2})$, the quantum envelpoing
affine Lie algebra, has been developed vigorously through statistical models,
and conformal field theories \cite{dri}.
{}From the progress of them, we have understood that the representation theory
of $q$-deformed universal enveloping algebras had also a deep connection with
many other fields of mathematics which were thought to be disconnected before.

In the study of integrable systems \vi\ algebra plays an essential part.
There have been some attempts to construct $q$-deformed version of \vi\
algebra \cite{poly,hiro,s1}.
None of the efforts were quite successful and the problem is left open.
This situation should be contrasted with one of the Kac-Moody case, in which
the $q$-deformed version has been studied in detail.

Some years ago, we proposed an algebra which is supposed to be a candidate
of the $q$-deformation of Virasoro algebra \cite{hiro,s1}:
\begin{equation}
[L_{mn},L_{m'n'}]=C_{m',n'}^{m,n}L_{m+m',n+n'}+C_{m',-n'}^{m,n}L_{m+m',n-n'}
\end{equation}
\bd
C_{m',n'}^{m,n}=
{{\left[{nm'-n'm \over 2}\right] [n+n']} \over {[n] [n']}}
\ed
where $[x] =(q^{x}-q^{-x})/(q-q^{-1})$. This is an infinite dimensional
Lie algebra which reduces to the ordinary
Virasoro in the $q\rightarrow1$ limit;
\begin{equation}
[L_{m},L_{m'}]=(m'-m)L_{m+m'}\ ,
\end{equation}
irrespective to the values of $n$ and $n'$.
If we restrict the generators to $L_{1,1}$, $L_{-1,1}$ and $L_{0,2}$, they
form the quantum group $U_{q}(sl_{2})$.
In this sense, (1) is a Lie algebra which includes $U_{q}(sl_{2})$ as a
``subalgebra''.
The central extension of (1) was dicussed in ref \cite{sato}.
Some years ago Chaichian and Pre${\rm \check{s}}$najdar showed that
the central extended version of the algebra (1)
can be represented by $q$-deformed fermionic fields starting from the quantum
Heisenberg algebra \cite{chai}.
But strictly speaking, it is not a so-called quantum group because it also
contains trivial Hopf structure.
Then, what is this algebra?
This is our problem discussed in this article.

We like to clarify the meaning of this algebra from the view point of
integrable deformation of Lie algebras.
Such deformation takes place in our case through discretization of differential
operators in the representation of \vi\ algebra.
The new algebras are reduced to the algebra of Chaichian et al. and ours under
 some conditions.
It also enables us to understand various steps which connect the Moyal algebra
and ours.
It will be shown that the discretized \vi\ algebra can be represented in terms
of free fields.
The symmetry of the algebra will be shown to correspond to the statistics of
the fields.
We will also discuss the central extension of our algebra.

\section{The $q$-discretization of Virasoro algebra}

To begin with let us notice the differential representation of $L_{mn}$
in (1), i.e.
\be
L_{mn}=z^{m}{q^{n(z\partial_{z}+\Delta_{m})}-q^{-n(z\partial_{z}+\Delta_{m})}
\over q^{n}-q^{-n}}
\equiv z^{m} \Bigl[ z\partial_{z}+\Delta_{m} \Bigr]_{n}
\ee
\bd
\Delta_{m}= {m \over 2}+\delta \hspace{.2in}
\  \forall \ \delta\ ,
\ed
where $[x]_{n} \equiv [x]/[n]$.
Note that $L_{1,1}$, $L_{-1,1}$ and $L_{0,2}$ are $U_{q}(sl_{2})$ generators
and form the quantum group.
We see that the main part of the operator is
\begin{equation}
{q^{n(z\partial_z+\Delta)}-q^{-n(z\partial_z+\Delta) } \over {q^{n}-q^{-n}}} ,
\end{equation}
which is nothing but a $q$-difference analog of the differential operator
 $z\partial_z+\Delta_{m}$.
So the algebra generated by $L_{mn}$ is a discrete version of
the corresponding Lie algebra.
We will discuss about algebras of this type and which reduce to the \vi\
algebra as $q$ goes to 1.

Firstly we pay attention to the fact that $L_{mn}$ consists of a shift
 operator  $l_{mn}^{\al}=z^{m}q^{nz\partial_{z}+\al}$, which
constitutes an algebra among themselves:
\be
[\ l_{mn}^{\al}\ ,\ l_{m'n'}^{\al '}\ ]=(q^{m'n}-q^{mn'})\
l_{m+m',n+n'}^{\al+\al '}.
\lb{fu}
\ee
We can link up this generator with other object, called the Moyal bracket
algebra.
For a function such as
\be
f(x,y)=\lambda \s_{m,n} f_{mn}(q)\  e^{mx}\ e^{-ny},
\ee
we define $K_{f}$ by
\be
K_{f} \equiv {1 \over 2}
 f(x+2\lambda \partial_{y}\ ,\ y-2\lambda \partial_{x})\ .
\ee
Then we find $K_{f}$ ($\forall f$) satisfies
\be
[K_{f},K_{g}]=K_{(\sinh 2\lambda \lbrace f,g \rbrace)}\ ,
\ee
\be
\sinh 2\lambda \{ f,g \} \equiv
\left. \sinh \Biggl[ 2\lambda \left(
    {{\pa} \over {\pa {x}_{f}}} {{\pa} \over {\pa {y}_{g}}}
     -{{\pa} \over {\pa {y}_{f}}} {{\pa} \over {\pa {x}_{g}}} \right)
\Biggr] f({x}_{f},{y}_{f})\ g({x}_{g},{y}_{g})
\right|_{{}^{x_{f}=x_{g}=x}_{y_{f}=y_{g}=y}}\ .
\ee
If we put $x=\ln z$, $y=\ln w$ and $\lambda=\ln q/2$ in (7), $K_{f}$ becomes
\be
 {\lambda \over 2} \s_{mn} f_{mn}\ l_{mn}^{\al}(z)\ l_{-n\ m}^{-\al}(w)\ .
\ee
This fact shows that
$l_{mn}^{\al}$ is
considered as a block of the generators $K_{f}$.
Eq.(8) is nothing but the Moyal bracket algebra.
Therefore the algebra (5) constitutes a part of the Moyal algebra \cite{f-z}.

The Moyal bracket was first introduced as a quantum mechanical analog or an
extension
of the Poisson bracket for the purpose of denoting statistical distributions on
quantum phase space \cite{moy}.
In this sense, our prescription of replacing the differential operators in
the \vi\ algebra by the difference operators provides an alternative
``quantization'' scheme of the $x$--plane.

Keeping this in mind we try to make certain combinations of $l_{mn}^{\al}$
which form a closed algebra and reduce to the \vi\ algebra in the $q\ra1$
limit.
We notice that
\be
{{l_{mn}^{\al}- l_{m\ -n}^{-\al}} \over {q-q^{-1}}}=
     z^{m}\Bigl[nz\pa_{z}+\al \Bigr]
\ee
do not close an algebra in general. But the following particular combinations
\be
 {l_{m\ j+r}^{jm \over 2}- l_{m\ -j-r}^{-{jm \over 2}} \over {q-q^{-1}}}
  \pm {l_{m\ j-r}^{jm \over 2}- l_{m\ -j+r}^{-{jm \over 2}} \over {q-q^{-1}}}
\ee
do for any fixed $r$; namely, if we rewrite this combination in more
convenient way as
\be
{\cal L}_{m}^{(j,r;\pm)}= z^{m}
{\left[ j \left(z\pa_{z} +{m \over 2} \right) \right] _{\mp} \over [j]_{\mp}}
         { \Bigl[ r z\pa_{z} \Bigr]_{\pm} \over [r]_{\pm}}
  \equiv  z^{m}
\left[ z\pa_{z} +{m \over 2}  \right] _{j;\mp}
         \Bigl[ z\pa_{z} \Bigr]_{r;\pm} \ ,
\lb{dvi}
\ee
where $[\ ]_{\pm}$ are defined as follows;
\bd
 [x]_{+}\equiv{{q^{x}+q^{-x}} \over 2}\ ,\
 [x]_{-}\equiv[x]\ ,
\ed
then we have
\bea
\Bigl[ {\cal L}_{m}^{(j,r;\pm)}, {\cal L}_{m'}^{(j',r;\pm)} \Bigr]
&=&
C(_{m'\ j'+ r}^{m\ \ j+r})_{\pm}\ {\cal L}_{m+m'}^{(j+j'+r,r;\pm)}+
C(_{m'\ j'-r}^{m\ \ j-r})_{\pm}\ {\cal L}_{m+m'}^{(j+j'-r,r;\pm)}
\nonumber \\
&& \hspace{.1in} + C(_{m'\ -j'+r}^{m\ \ \  j+r})_{\pm}
\ {\cal L}_{m+m'}^{(j-j'+r,r;\pm)}
+C(_{m'\ -j'-r}^{m\ \ \ j-r})_{\pm}\ {\cal L}_{m+m'}^{(j-j'-r,r;\pm)}
\lb{dv2}
\eea
for each value of $r$.
The double signs on both sides correspond each other.
The structure constant $C$ is defined as
\be
C(_{m'\ j'+r}^{m\ \ j+r})_{\pm} \equiv
{{\Bigl[ {{(j+r)m'-(j'+r)m} \over 2} \Bigr]_{-}\  [j+j'+r]_{\mp}} \over
{2\  [j]_{\mp}\  [j']_{\mp}\  [r]_{\pm}}}\ .
\ee
These algebras contain $L_{mn}$ algebra of (1) and the algebra of Chaichian
et al. Actually, the generators of the former can be identified with
${\cal L}_{m}^{(j,0;+)}$, and
the latter with ${\cal L}_{m}^{(j,1;-)}$, respectively.
In the limit of $q\ra1$, ${\cal L}_{m}^{(j,r;\pm)}$ reduce such as
\be
{\cal L}_{m}^{(j,r;+)} \ra  z^{m} \Bigl(z\pa_{z}+  {m \over 2} \Bigr) \equiv
L_{m}^{(+)}, \hspace{.3in}
{\cal L}_{m}^{(j,r;-)} \ra  z^{m+1} \pa_{z} \equiv L_{m}^{(-)},
\ee
respectively.
These $L_{m}^{(\pm)}$ satisfy (2), that is they are \vi\  generators.
Thus we get $q$-discretized \vi\ algebras, which we call \dv\ algebras
(or \dv\ for short).

More generally, if we sum over $j$ and $r$ under the proper weight, i.e. we
define
\be
K_{m}^{(f;\pm)}= z^{m} \s_{j,r} \varepsilon_{r}^{(\pm)}\ F_{m\ j}^{(\pm)}
\Bigl[ z\partial_{z}+{m \over 2} \Bigr]_{j;\mp}
             \Bigl[ z\partial_{z} \Bigr]_{r;\pm},
\lb{mm}
\ee
these operators also form a closed algebra:
\bea
\Bigl[K_{m}^{(f;\pm)}\ ,\ K_{m'}^{(g;\pm)}\Bigr]
&=&  z^{m+m'} \sum_{J,r}  \varepsilon_{r}^{(\pm)}\ H_{m+m'\ J}^{(\pm)}
         \biggl[  z\partial_{z}+{{m+m'} \over 2} \biggr]_{J;\mp}
               \Bigl[ z\partial_{z} \Bigr]_{r;\pm} \nonumber \\
&=&  K_{m+m'}^{(h;\pm)}
\eea
where
\bea
H_{m+m'\ J}^{(\pm)} &=& \sum_{j+j'+s=J} \varepsilon_{s}^{(\pm)}\ F_{m\
j}^{(\pm)}\
G_{m'\ j'}^{(\pm)}\
C(_{m'\ j'+s}^{m\ \ j+s})_{\pm}
     + \sum_{j+j'-s=J} \varepsilon_{s}^{(\pm)}\ F_{m\ j}^{(\pm)}\ G_{m'\
j'}^{(\pm)}\
 C(_{m'\ j'-s}^{m\ \ j-s})_{\pm}
\nonumber \\
&&
     +\sum_{j-j'+s=J} \varepsilon_{s}^{(\pm)}\ F_{m\ j}^{(\pm)}\  G_{m'\
j'}^{(\pm)}\
C(_{m'\ -j'+s}^{m\ \ \ j+s})_{\pm}
      +\sum_{j-j'-s=J} \varepsilon_{s}^{(\pm)}\ F_{m\ j}^{(\pm)}\ G_{m'\
j'}^{(\pm)}\
C(_{m'\ -j'-s}^{m\ \ \ j-s})_{\pm} \nonumber \\
&=&4\  \s_{j+j'+s=J} \varepsilon_{s}^{(\pm)}\ F_{m\ j}^{(\pm)}\ G_{m'\
j'}^{(\pm)}\
C(_{m'\ j'+s}^{m\ \ j+s})_{\pm}\ .
\lb{gen}
\eea
The last line of (\rf{gen}) is derived by setting
$\varepsilon_{s}=\varepsilon_{-s}$ and $F_{j}=F_{-j}\ (G_{j}=G_{-j})$.

The Moyal deformation has also been discussed as
a deformation of $w_{\infty}$ algebra within the context of the
area preserving maps on a torus \cite{pop,winf}.
 This is natural from our point of view, because the $w_{\infty}$ algebra
contains Virasoro as a subalgebra ($N=2$) \cite{bak}.
Hence $D$--\vi \  might have some connections with deformations of
$w_{\infty}$ algebra.

\section{The free field representation of $D$--Virasoro}

In this chapter, we consider the free field representation of \dv\ in the
bilinear form of free fields.
The difference between ordinary \vi\  and \dv\ is that the
latter has new indices $j$, $r$ and $\pm$.
Though $j$ is different for each generator, $r$ and $\pm$
are common for all.
This means that there are different types of \dv\ algebras specified by
$r$ and $\pm$ which are independent each other.
Hence it is natural to label the fields by them.

We first define $\wh{\cal L}_{m}$ by
\begin{equation}
{\widehat{\cal L}}_{m}^{(j,r)}={1 \over 2}\sum_{k}
A_{k\ m-k}^{(j)} (q)
\ \alpha_{m-k}^{(r)}\   \alpha_{k}^{(r)},
\lb{bil}
\end{equation}
Next, we assume that the free fields $\al_{k}^{(r)}$ are bosonic,
i.e. they satisfy
\be
{[\alpha_{k}^{(r)},\alpha_{k'}^{(r)}]} \equiv  D_{k}^{(r)}(q)\
\delta_{k+k',0}\ ,
\lb{com}
\ee
where $D_{k}^{(r)}(q)$ is a function of $q$ which is to be determined later.
Accordingly, $A_{k\ m-k}^{(j)}$ and $D_{k}^{(r)}$ must satisfy the following
symmetry conditions;
\be
 A_{k\ m-k}^{(j)}=A_{m-k\ k}^{(j)} \ , \
 D_{k}^{(r)}=- D_{-k}^{(r)}\ .
\ee
so that ${\wh{\cal L}}_{m}^{(r)}$ is defined properly.
We now ask if this ${\wh{\cal L}}_{m}$ surely the bosonic free field
representation of the \dv.
To answer this let us calculate the commutator of ${\wh{\cal L}}_{m}$;
\bea
\Bigl[{\wh{\cal L}}_{m}^{(j,r)},{\wh{\cal L}}_{m'}^{(j',r)} \Bigr] &=&
- {1 \over 2}\sum_{k}
 \biggl( A_{k-m'\ m+m'-k}^{(j)}\ A_{k\ m'-k}^{(j')}\
              D_{m'-k}^{(r)} \nonumber \\
 && \hspace{.7in}  -A_{k\ m-k}^{(j)}\ A_{k-m\ m+m'-k}^{(j')}\
              D_{m-k}^{(r)} \biggr)\
\alpha_{m+m'-k}^{(r)}\ \alpha_{k}^{(r)} \ ,
\lb{34}
\eea
On the other hand, if we substitute ${\wh{\cal L}}_{m}$ to ${\cal L}_{m}$ and
calculate the r.h.s. of (\ref{eqn:dv2}), we find
\bea
&&{1 \over 2} \sum_{k}  \biggl(
      C(_{m'\ j'+r}^{m\ \ j+r})_{\pm}\  A_{k\ m+m'-k}^{(j+j'+r)}\
 +C(_{m'\ j'-r}^{m\ \ j-r})_{\pm}\ A_{k\ m+m'-k}^{(j+j'-r)}  \nonumber \\
&& \hspace{.4in}
   + C(_{m'\ -j'+r}^{m\ \ \ j+r})_{\pm}\  A_{k\ m+m'-k}^{(j-j'+r)}\
  + C(_{m'\ -j'-r}^{m\ \ \ j-r})_{\pm}\ A_{k\ m+m'-k}^{(j-j'-r)} \biggr)\
         \alpha_{m+m'-k}^{(r)}\ \alpha_{k}^{(r)}\ .
\lb{35}
\eea
Comparing (\rf{34}) and (\rf{35}), we find a simple solution for $A$ and $D$,
if we take the index ``$-$'', i.e. the structure constant $C_{-}$ in (\rf{35});
\be
A_{k\ m-k}^{(j)}=-\biggl[ {{2k-m} \over 2} \biggr]_{j;+}\ , \
D_{k}^{(r)}=[ k ]_{r;-}\ .
\label{eqn:36}
\ee
We cannot find such solution in ``+'' case.
This means that ${\wh{\cal L}}_{m}^{(j,r;-)}$ can be represented by the
use of bosonic operators only.

Then corresponding to ${\cal L}_{m}^{(j,r;+)}$, it is natural to assume that
they are
represented by the fermionic free fields.
Actually, if we define the anticommutation relation of them by
\be
[\alpha_{k}^{(r)},\alpha_{k'}^{(r)}]_{+} \equiv  D_{k}^{(r)}(q)\
\delta_{k+k',0}\ ,
\ee
where $[\ ,\ ]_{+}$ denotes the anticommutator,
then we can show that ${\wh{\cal L}}_{m}^{(j,r)}$ forms an algebra isomorphic
to (\ref{eqn:dv2}) with ``$+$'',
for the following $A$ and $D$;
\be
A_{k\ m-k}^{(j)}=-\biggl[ {{2k-m} \over 2} \biggr]_{j;-}\ , \
D_{k}^{(r)}=[ k ]_{r;+}\ \ .
\ee
{}From these results, we see that the index $\pm$ in (\ref{eqn:dvi})
comes from the difference of the
statistics of the free field representation of them.
If we take the limit of $q\ra1$, ${\wh{\cal L}}$ reproduces the original \vi\
generators represented by boson and fermion for $-$ and $+$, respectively.

\section{Central extension}

Though we have neglected the center of \dv\ in our discussion so far,
it is straightforward to incorporate it to our theory. Let us investigate
the bosonic case.
For later convenience,
 we change the notation and denote the bosonic field by use of the Hermite
conjugate operators
\be
\alpha_{k}^{(r)}=-iB_{k}^{(r)}(q)\  a_{k}^{(r)} \ \ ,\  \
\alpha_{-k}^{(r)}=i{\bar B}_{k}^{(r)}(q)\  a_{k}^{(r) \dagger} \ ,\
\ee
where $B_{k}({\bar B}_{k})$ is a function of $q$
 and tends to $\sqrt k$ in the
$q\rightarrow1$ limit.
The commutation relation (\rf{com}) becomes
\be
[a_{k}^{(r)},a_{k'}^{(r)\dagger}]
      ={D_{k}^{(r)} \over {B_{k}^{(r)}\ {\bar B}_{k}^{(r)}}}\ \delta_{k,k'}.
\ee
For the zero mode, we assume for the operators to satisfy the relation;
\begin{equation}
[x,p]=i\ {{q^{r}-q^{-r}} \over 2r {\ln q}} \equiv iD_{0}(q).
\end{equation}
(Hereafter, for simplicity, we will not write the index $r$ explicitly.)
In this notation, after the normal-ordering, ${\wh{\cal L}}$
is represented by
\bea
{\wh{\cal L}}_{m}&=& iA_{m\ 0}^{(j)}\ D_{0}^{-1}\ B_{m}\ p\ a_{m}
+{1 \over 2} \s_{k=1}^{m-1} A_{k\ m-k}^{(j)} \ B_{m-k} \ B_{k} \
a_{m-k}\ a_{k}
\nonumber \\
&& \hspace{1.8in} - \s_{k=m+1}^{\infty} A_{k\ m-k}^{(j)}\ {\bar B}_{k-m}\
B_{k}
\ a_{k-m}^{\dagger}\ a_{k}
 \\
{\widehat{\cal  L}}_{-m}&=& - iA_{m\ 0}^{(j)}\ D_{0}^{-1}\ {\bar B}_{m}\
p\ a_{m}^{\dagger}
+{1 \over 2} \s_{k=1}^{m-1} A_{k\ m-k}^{(j)}\ {\bar B}_{m-k}\
{\bar B}_{k}\ a_{m-k}^{\dagger}\ a_{k}^{\dagger}
\nonumber \\
&& \hspace{1.8in} - \s_{k=m+1}^{\infty}\ A_{k\ m-k}^{(j)}\ {\bar B}_{k}\
B_{k-m}
\ a_{k}^{\dagger}\ a_{k-m}
\\
{\wh{\cal L}}_{0}&=&- {D_{0}^{-2} \over {2\ [j]_{+}}}\  p^{2}
+ \s_{k=1}^{\infty} A_{k\ -k}^{(j)}\ {\bar B}_{k}\ B_{k}
     \  a_{k}^{\dagger}\ a_{k} \ .
\eea
for $m>0$.
The center can be obtained from the commutator
$[{\wh{\cal L}}_{m},{\wh{\cal L}}_{-m}]$ ($m\neq0,\pm1$) by taking the
normal-ordering of $\wh{\cal L}_{0}$;
\be
{1 \over 2} {\s_{k=1}^{m-1}} A_{k\ m-k}^{(j)}\ A_{k\ m-k}^{(j')}\
D_{k}\ D_{m-k}\ .
\ee
When we substitute $A$ and $D$ of (\ref{eqn:36}), it becomes
\be
{1 \over 2} {\s_{k=1}^{m-1}}
\biggl[ {{2k-m} \over 2} \biggr]_{j;+}
\biggl[ {{2k-m} \over 2} \biggr]_{j';+}
 [k]_{r;-} [m-k]_{r;-}\ .
\lb{center}
\ee
It is easy to check that this central extension satisfies Jacobi identity
and the center (\rf{center}) reduces to $m(m^{2}-1)/12$ in
$q \rightarrow 1$.
For the fermionic case, we can also get it in a similar manner.

\section{Discussions}

In this paper we have constructed the $q$-discretized \vi\ algebra,
in short, $D$--Virasoro.
Though we use the term ``$q$-disretization'', its full meaning has not been
clarified.
At the present stage, we can only say that the $q$-discretization means the
discretization of Riemann surfaces on which fields are defined.

On the other hand, our scheme is certainly ``quantization'' at the algebraic
level because it is originated from the Moyal bracket algebra
which was quantum mechanical analog of the classical Poisson algebra.
{}From this point of view, the motivation of the $q$-discretization is similar
to the one of $q$-deformation \cite{icm}.
But $D$--Virasoro is not a nontrivial quantum group because it contains the
trivial Hopf algebra even at $q\not=1$.
(In ref.\cite{hiro,chai} it is shown that their algebras have
the non-trivial Hopf structure. Then we expect $D$--Virasoro also has such
 one.)
As the $q$-deformation of Virasoro has not been discovered yet,
we need make clear the relation between these two different schemes of
``quantizing'' algebras.
$D$--Virasoro may give some informations for the study of it.
The break through seems to be in the affine Kac-Moody algebra, too.
The research of $U_{q}({\widehat {sl_{2}}})$ has been much progressed
as mentioned above \cite{inf,dri}.
If we get its $q$-discretized form, we must be able to show the
connection between the $q$-discretization and the $q$-deformation.

Following the introduction of \dv, we represented it by the use of the free
fields.
It is nothing but a $q$-discretized Sugawara construction.
There it was shown that the bosonic and fermionic representations could be
understood in a unified way.
{}From this fact we are led to the concept of supersymmetric algebras.
If we consider the supersymmetric Riemann surfaces, two different types of
\dv\ might be represented by a superalgebra \cite{chai}.
Moreover, use of such representation enables us to discuss deformation of
integrable field theories.
At present it is uncertain whether such deformation preserves integrability.
But there are some indications which seem to
support the integrability of the $q$-discretized models.
For example, in string theory, a braiding of
a kind of deformed strings is shown to yield a Yang-Baxter
equation \cite{s1,c-s}.
This problem is also connected to the representation theory of \dv.
It is expected that the investigation of the representation theory reveals
the physical meanings of $q$-discretization.

\vskip .5cm
We would like to thank Chikara Itoh and Katsumi Shigura for useful
calculation.

\end{document}